\documentclass[sigconf]{acmart}

\AtBeginDocument{%
  }

\setcopyright{acmlicensed}
\copyrightyear{2018}
\acmYear{2018}
\acmDOI{XXXXXXX.XXXXXXX}

\acmConference[Conference acronym 'XX]{Make sure to enter the correct
  conference title from your rights confirmation emai}{June 03--05,
  2018}{Woodstock, NY}
\acmISBN{978-1-4503-XXXX-X/18/06}

\usepackage{graphicx}
\usepackage{wrapfig}
\usepackage[T1]{fontenc}
\usepackage{amsmath}  
\usepackage{multirow}
\usepackage{tabularx}
\newcommand{\x}[1]{{\leavevmode\color{black}{#1}}}

\usepackage{soul}
\usepackage{adjustbox}
\usepackage{subcaption}

\begin{document}

\title{Exploring Teenagers’ Trust in Al Chatbots: An Empirical Study of Chinese Middle-School Students}

\author{Siyu Qiu}
\authornote{These authors contributed equally to this work.}
\orcid{}
\email{23300130041@m.fudan.edu.cn}
\affiliation{%
  \institution{Fudan University}
  \city{Shanghai}
  \country{China}
}

\author{Anqi Lin\textsuperscript{*}}
\orcid{}
\email{lynn027ovo@gmail.com}
\affiliation{%
  \institution{Fudan University}
  \city{Shanghai}
  \country{China}
}

\author{Shiya Wang\textsuperscript{*}}
\orcid{}
\email{jessie181010@gmail.com}
\affiliation{%
  \institution{Fudan University}
  \city{Shanghai}
  \country{China}
}

\author{Xingyu Lan}
\email{xingyulan96@gmail.com}
\authornote{Xingyu Lan is the corresponding author. She is with Fudan University and a member of the Research Group of Computational and AI Communication at Institute for Global Communications and Integrated Media. }
\affiliation{%
  \institution{Fudan University}
  \city{Shanghai}
  \country{China}
}

\newcommand{\etal}{et~al.~} 
\newcommand{\ie}{i.e.,~}
\newcommand{\eg}{e.g.,~}

\renewcommand{\shortauthors}{Lan et al.}

\renewcommand{\sectionautorefname}{Section}
\renewcommand{\subsectionautorefname}{Section}
\renewcommand{\subsubsectionautorefname}{Section}

\begin{abstract}
 Chatbots have become increasingly prevalent. A growing body of research focused on the issue of human trust in AI. However, most existing user studies are conducted primarily with adult groups, overlooking teenagers who are also engaging more frequently with AI technologies. Based on previous theories about teenage education and psychology, this study investigates the correlation between teenagers’ psychological characteristics and their trust in AI chatbots, examining four key variables: AI literacy, ego identity, social anxiety, and psychological resilience. We adopted a mixed-methods approach, combining an online survey with semi-structured interviews. Our findings reveal that psychological resilience is a significant positive predictor of trust in AI, and that age significantly moderates the relationship between social anxiety and trust. The interviews further suggest that teenagers generally report relatively high levels of trust in AI, tend to overestimate their AI literacy, and are influenced by external factors such as social media.
\end{abstract}




\maketitle

\section{Introduction}

The widespread integration of artificial intelligence (AI) into daily life has captured a vast and diverse user base, including teenagers. As teenagers undergo significant physiological and social transitions~\cite{sarah2014adolescence}, these developmental characteristics may make them particularly vulnerable to forming relationships with and placing trust in AI systems.
The potentially severe consequences of this vulnerability are underscored by recent reports of tragic cases where teen suicides were linked to extensive interactions with AI chatbots (\eg \cite{Kashmir,LauraReiley}).
These incidents highlight a growing problem: many teenagers place significant trust in AI, disclosing sensitive personal and emotional information despite the lack of genuine relational reciprocity. While AI is not the direct cause of underlying mental health issues such as depression in these cases, its role as a preferred confidant—potentially displacing interpersonal relationships—may exacerbate existing risks. This dynamic necessitates urgent scholarly inquiry into teen trust in AI, particularly among psychologically vulnerable teenagers.

In academia, human–AI trust has already become a mainstream research topic. As early as 1994, Nass~\etal proposed the ``Computers Are Social Actors'' (CASA) paradigm, positing that even experienced users instinctively apply social rules when interacting with computers~\cite{nass1994computers}. They further argued that as long as machines display sufficient human-like cues, people will naturally equate computers with human beings~\cite{reeves1996media}. Building on this foundation, research on emerging AI chatbots has identified more specific trust factors. For instance, Go and Sundar~\cite{go2019humanizing} found that message interactivity is a key factor in humanizing chatbots, and that identity cues, by shaping users' expectations, influence their trust in and evaluation of such agents. Mostafa and Kasamani~\cite{mostafa2022antecedents} found that in addition to performance expectancy, compatibility, perceived ease of use, and social influence all significantly enhance consumers’ initial trust in chatbots. 

However, existing user studies have mainly focused on adults~\cite{glikson2020human,zhang2023trust}, while under-exploring the perspectives of teenagers, who are also frequent users of AI. According to Brandtzaeg~\etal~\cite{brandtzaeg2025emerging}, social AI services such as ChatGPT are becoming increasingly integrated into teenagers’ daily lives. Data show that four out of five (79\%) online teenagers in the United Kingdom (ages 13–17)~\cite{ofcom_onlinenation_2024} and 51\% of American teenagers (ages 14–22)~\cite{hopelab_generativeAI_2024} have used generative AI tools at least once. In China, teenagers (ages 6–19) account for 21.1\% of all generative AI users~\cite{CNNIC2025}. Specifically, according to a 2025 survey by the Pew Research Center of American teenagers aged 13–17, 79\% reported having heard of chatbots such as ChatGPT, and 26\% reported using it for school assignments~\cite{sidoti_etal_chatgpt_2025}.

Moreover, since teenagers are at a critical stage of forming personal values and developing abilities, how they use AI technologies appropriately may significantly affect their personal growth~\cite{piombo2025emotional}. As one of the few studies focusing on teenagers’ trust in AI, Piombo et al.~\cite{piombo2025emotional} investigated generational differences and influencing factors in teenagers’ AI use and trust, emphasizing the importance of family relationships and emotional competence in healthy AI use. Lee et al.~\cite{lee2025understanding}, using the Design Fiction method, identified the complexity of teenagers’ health conditions, the lack of AI empathy, and their prior experiences with AI as key factors shaping teenagers’ trust in AI. However, studies like these often examine trust through external factors such as socio-demographic variables or user experience, paying less attention to the internal, developmental psychological characteristics of teenagers themselves.

To address this gap, this work employs a sample of Chinese middle school students to examine how four key developmental attributes—AI literacy, ego identity, social anxiety, and psychological resilience—shape teenagers' trust in AI chatbots. Grounded in established psychological and educational theories (see \autoref{sec:related} for more details), these factors are hypothesized to be potentially salient in forming trust during adolescence. Specifically, we adopted a mixed-methods approach, starting with an online survey (N = 153) to test the hypothesized correlations between these factors and trust. Subsequently, semi-structured interviews were conducted to provide complementary qualitative insights, particularly for relationships that were not statistically significant. 

In a word, we found that psychological resilience has a significant positive effect on trust, age significantly moderates the relationship between social anxiety and trust, gender differences significantly influence trust, and teenagers generally overestimate their AI literacy. 
\x{Meanwhile, in our study, the sample of middle school students from China demonstrated a strong tendency to treat AI primarily as a tool. They mainly used AI chatbots for functional purposes such as information retrieval and writing refinement to support their academic and exam-oriented needs, rather than regarding AI as genuine friends, companions, or life partners. In other words, it is important to clarify that the conclusions drawn in this study are in fact highly focused on teenagers' human-AI trust in learning contexts; also, we did not observe the strong emotional trust (or even overtrust) in our sample that has been described in some media reports.} 
By focusing on teenagers' intrinsic psychological landscape, this study contributes empirical depth to theories of AI trust and offers practical guidance for educators, developers, and policymakers to foster healthier AI interactions for teenagers.

\section{Related work}
\label{sec:related}

Below, we review previous work about human-AI trust and introduce how we construct our hypotheses based on relevant literature.

\subsection{Human-AI Trust}
\label{ssec:Trust}

In classic discussions of interpersonal trust, \textit{trust} is defined as a state in which \x{``one party is willing to expose themselves to the risk of the other's behavior, expecting the other to perform an action that is crucial to them, regardless of whether they can monitor or control that other party''} ~\cite{mayer1995integrative}. 
Luhmann conceptualized trust as a crucial social heuristic that reduces complexity in situations of uncertainty~\cite{luhmann2018trust}.
Mayer et al.’s integrative model posits that perceptions of a trustee’s ability, benevolence, and integrity are central to trust formation~\cite{mayer1995integrative}.

In the context of human-computer interaction (HCI), researchers have conducted a series of experiments in the last century, finding that trust mechanisms were important factors in the use and adoption of software and applications (\eg ~\cite{viega2001trust,massa2005pagererank}), and human-computer trust was manageable~\cite {khare1997weaving,grandison2009survey}. In 2004, Lee and See~\cite{lee2004trust} defined human-machine trust as ``the attitude that an agent will help achieve an individual’s goals in a situation characterized by uncertainty and vulnerability''.

This foundational conceptualization has further spurred extensive research on user trust toward various technologies, such as robots~\cite{hancock2011meta,campagna2025systematic}, voice assistants~\cite{choung2023trust}, and autonomous vehicles~\cite{jing2020determinants}.
Recently, the advent of generative AI has introduced both new opportunities and challenges~\cite{yu2024exploring}. Unlike traditional technologies, AI exhibits distinct characteristics in its performance, attributes, and purposes, making trust in AI a unique phenomenon~\cite{siau2018building,lan2025imagining}. 

Recent empirical studies have provided valuable insights into the factors shaping human-AI trust across different contexts.
Choung et al.~\cite{choung2023trust} analyzed AI voice assistants and classified AI trust into human-like trust and functionality trust, incorporating it into the Technology Acceptance Model (TAM) to confirm that the reliability and functionality of AI are central determinants of users' adoption~\cite{choung2023trust}. Baughan et al.~\cite{baughan2023mixed} further conducted a mixed-methods study on voice assistant failures, finding that different failure types differentially affect user trust, with overly capturing users' input causing greater trust erosion than ambiguity errors. Pitts et al.~\cite{pitts2025understanding} explored AI trust in education, and found that trust in AI was blending elements of interpersonal trust and technological trust, with anthropomorphism making the borders of these two trust nuanced~\cite{pitts2025understanding}. A large-scale study by Li et al. ~\cite{li2025human} examined AI search systems, demonstrating that design features such as citations, social feedback, and uncertainty cues significantly shaped user trust.

While previous research has made significant progress in defining and evaluating human–AI trust, the literature has predominantly focused on the characteristics of AI as a \textit{trustee}—such as its ability, transparency, or anthropomorphism—in shaping trust. In doing so, it has largely treated the \textit{trustor} as a homogeneous entity, paying insufficient attention to individual differences in cognition and psychology. Moreover, the studies have almost exclusively relied on adult samples, largely overlooking the growing demographic of teenage AI users. 

\x{\subsection{Factors Influencing Human-AI Trust}}

To address these gaps, this study focuses specifically on teenagers, investigating how four of their key cognitive and psychological characteristics (introduced individually in the following sections) influence trust formation in AI.

\subsubsection{AI Literacy}
\label{ssec:AI Literacy and Human-AI Trust}

\x{The concept of  \textit{literacy} has evolved beyond the ability to express and communicate through written language to encompass more complex learning processes and is now used to define skill sets in various disciplines, such as digital literacy and computational literacy~\cite{long2020ai}.}
There is currently no unified definition of AI literacy in academia, but it generally includes both cognitive and practical abilities related to AI. \x{For instance, Long and Magerko~\cite{long2020ai} defined AI literacy as a range of abilities to critically assess AI technologies, interact effectively with them, and apply AI as a tool across various contexts.} Ng et al.~\cite{ng2021conceptualizing} further reinforced this definition by suggesting that AI literacy can be cultivated through four dimensions: understanding and comprehension, use and application, evaluation and creation, and ethical issues. 
\x{Wagner~\cite{wagner2021economic} found that AI literacy enables individuals to make informed decisions, engage in discussions about AI’s role in society, and adapt to AI-driven advancements. A high level of AI literacy allows individuals to more accurately perceive the boundaries and operational principles of AI, assess its capability and trustworthiness, and form ``appropriate trust''~\cite{lee2004trust} that aligns with AI's capabilities and helps avoid inappropriate reliance.}

In the context of AI being integrated into school education, it has become particularly important to foster AI literacy in teenagers through interdisciplinary and competency-based approaches~\cite{casal2023ai,heintz2021three}. AI-literacy is also considered a core competency to prepare teenagers to be AI developers, which enhances teenagers' proactive engagement with AI~\cite{su142315620}. In a word, AI literacy may help teenagers form an accurate perception of AI's capabilities, which in turn influences the level of trust they place in AI. Based on this, this study hypothesizes:

\textit{\textbf{H1}: Teenagers’ AI-literacy level is positively correlated with their trust in AI chatbots.}

\subsubsection{Ego Identity}
\label{ssec:Ego Identity and Human-AI Trust}

Ego identity, as a core concept in personality development, was first systematically proposed by Erikson~\cite{erikson1956problem}. It encompasses an individual's intrinsic continuity in the sense of ``who I am'' and the connection between the individual and specific groups in terms of values, history, or ideals.
Building on Erikson's framework, Marcia ~\cite{marcia2012ego} operationalized ego identity into two key variables: exploration and commitment. The interplay between these variables gives rise to four identity statuses: achievement, moratorium, foreclosure, and diffusion~\cite{marcia2012ego}. These statuses reflect different psychological characteristics of teenagers at various stages of identity formation and are linked to psychosocial issues and well-being~\cite{crocetti2009anxiety}. 
Erikson posited a key conflict in teenagers between achievement identity (a mature state forged through exploration and commitment) and diffusion identity (characterized by confusion, avoidance of commitment, and difficulties in forming intimate relationships)~\cite{erikson1956problem}. A stable, achieved identity allows individuals to establish clear personal boundaries and engage with the world from a secure self-concept~\cite{meeus1999patterns}. 

We posit that this extends to interactions with AI. Chatbots, designed to be user-friendly and empathetic, offer a low-risk environment for social interaction~\cite{zhai2022systematic}. \x{This very context makes it a pertinent domain for exercising ego identity: the act of forming trust—a social judgment—in a novel entity like AI mirrors the process of forming commitments and beliefs central to identity development. This view is corroborated by existing studies which have indicated a possible influence of ego identity in AI use (\eg ~\cite{rodriguez2025artificial,yao2025connecting}).}.
Therefore, we hypothesize that teenagers with an achieved identity are better equipped to interact with AI based on a rational understanding of its capabilities and limitations, \x{which fosters greater confidence in relying on it and manifests as higher trust. In contrast, those with a diffuse identity, lacking a stable basis for judgment, may approach the technology with greater uncertainty and skepticism, resulting in lower reliance.}
Thus, this study hypothesizes:

\textit{\textbf{H2}: Teenagers’ ego identity level is positively correlated with their trust in AI chatbots.}

\subsubsection{Social Anxiety}
\label{ssec:Social Anxiety and Human-AI Trust}

Social anxiety can be understood as the fear, worry, and avoidance emotions an individual experiences in the face of real or anticipated social situations. It includes three basic dimensions: fear of negative evaluation (FNE), social avoidance and distress in general situations (SAD-General), and social avoidance and distress in new situations (SAD-New)~\cite{la1998social}. Adolescence is a critical period for the development of social relationships, and social anxiety significantly impedes and suppresses peer interactions and the development of friendships~\cite{xin2022changes}.

While the Social Compensation Hypothesis suggests that socially anxious individuals may turn to online human interactions due to the reduced pressure of non-verbal cues and the controllability of the environment~\cite{valkenburg2007preadolescents,schouten2007precursors}, this behavior is ultimately a risk-avoidance strategy. It is motivated more by escaping the uncertainties of offline interaction than by a drive to build genuine trust, and it can lead to negative outcomes like increased loneliness and internet addiction~\cite{dong2024effect,bonetti2010relationship}. 
Interacting with AI chatbots presents a distinct and potentially more challenging scenario for socially anxious individuals. Unlike controlled online interactions with humans, trust in AI requires entrusting personal information and emotions to a ``black box'' system whose decision-making process is opaque and whose responses can be erroneous or unpredictable. This introduces a fundamental lack of controllability and predictability—precisely the factors that socially anxious individuals seek to maximize.
Therefore, although AI might seem like a safe alternative, its inherent uncertainty may be magnified through the lens of social anxiety, because AI may fail to meet their deep need for high-quality, authentic social interactions~\cite{rodriguez2025artificial}. 
Based on this, we hypothesize:

\textit{\textbf{H3}: Teenagers' social anxiety level is negatively correlated with their trust in AI chatbots.}

\subsubsection{Psychological Resilience}
\label{ssec:Psychological Resilience and Human-AI Trust}
Psychological resilience refers to an individual's ability to maintain adaptability and recover balance in the face of significant adversity ~\cite{luthar2000construct,masten2001ordinary}. Individuals with high psychological resilience usually possess greater autonomy, responsiveness, and self-efficacy~\cite{gabrielli2022promoting}. It has also been found that psychological resilience interventions can enhance teenagers’ individual skills and resources~\cite{kallianta2021stress}.
Relating to the topic of this work, Xiao et al.~\cite{xiao2019factors} found that psychological resilience impacts how users interact with AI chatbots and their purposes for doing so; users with high psychological resilience exhibit stronger social skills, efficiency, and autonomy, allowing them to use technology with greater intentionality. 

In research on teenagers, psychological resilience has received growing attention. As teenagers are exposed to increasing volumes of online information and digital media, their psychological resilience is both challenged and potentially enhanced~\cite{jack2019educational,meng2025human,mcgrew2018can}.
For example, Meng et al.~\cite{meng2025human} found that teenagers' ability to use technology can be improved through psychological resilience, serving as an intermediary factor that enhances children's school readiness; they also observed that psychological resilience can help teenagers better adjust and cope when faced with technological problems. 
Mcgrew~\etal~\cite{mcgrew2018can} found that without sufficient psychological resilience, teenagers may struggle to accurately assess information and form a reasonable understanding of the advantages and disadvantages of technology. 
Similarly, psychological resilience may also influence teenagers' perceptions of AI's capabilities, which in turn affects their trust in it.
Thus, this study hypothesizes:

\textit{\textbf{H4}: Teenagers' psychological resilience level is positively correlated with their trust in AI chatbots.}

\section{Questionnaire Survey of Middle-School Students}
\label{sec:Questionnaire}

In this section, we present a questionnaire survey conducted among Chinese teenagers.

\subsection{Operational Definitions and Measurement of Variables}

To measure the core variables of this study, we adopted standardized scales or multiple-choice items that have been widely validated in prior literature. 
Below, we introduce the measured metrics one by one.

\paragraph{AI Literacy}

This study used the AI Literacy Questionnaire developed by Ng et al. ~\cite{ng2024design} to measure teenagers’ artificial intelligence literacy. The scale provides a comprehensive assessment of students’ AI literacy across four dimensions: affective  (\eg ``Artificial intelligence is closely related to my daily life.''), behavioral (\eg ``I believe I can master the knowledge and skills related to AI.''), cognitive (\eg ``I can mainly utilize AI to solve problems.''), and ethical  (\eg ``I am aware of the potential risks that the overuse of AI poses to humanity.''). The above items were evaluated using a five-point Likert scale (1 = Not at all, 5 = Strongly agree).

\paragraph{Ego Identity}

To measure ego identity, this study employed the Ego Identity Scale developed by Tan et al.~\cite{tan1977short}. The scale comprises 12 items in a forced-choice format, each presenting two statements that reflect the conflict between identity exploration and role confusion as defined by Erikson’s psychosocial theory. For example, respondents need to choose between options such as ``I enjoy actively participating in various clubs and group activities'' and ``I prefer focusing on hobbies that I can pursue at my own pace and on my own schedule''. 

\paragraph{Social Anxiety}

We used the Social Anxiety Scale for Adolescents (SAS-A) developed by La Greca and Lopez~\cite{la1998social} to assess the level of social anxiety. The scale comprises three core dimensions: Fear of Negative Evaluation (FNE, \eg ``I worry about what others say about me.''), Social Avoidance and Distress in new situations (SAD-New, \eg ``I get nervous when I meet new people.''), and Social Avoidance and Distress in general situations (SAD-General, \eg ``It's hard for me to ask others to do things with me.''). The above items were evaluated using a five-point Likert scale (1 = Not at all, 5 = Strongly agree).

\paragraph{Psychological Resilience}

This study employed the 10-item Connor–Davidson Resilience Scale (CD-RISC), validated and refined by Campbell-Sills and Stein~\cite{campbell2007psychometric}, to measure psychological resilience. The scale is designed to evaluate individuals’ capacity to adapt to and recover from stress, trauma, and adversity (\eg ``Can stay focused under pressure''). The above items were evaluated using a five-point Likert scale (1 = Not at all, 5 = Strongly agree).

\paragraph{Trust}

We used the 12-item Human-Computer Trust Scale (HCTS) developed by Gulati et al.~\cite{gulati2019design}, which assesses key dimensions including perceived risk (\eg ``I believe that there could be negative consequences when using chatbots''), benevolence (\eg ``I believe that chatbots will act in my best interest''), competence (\eg ``I think that chatbots performs its role very well'') and reciprocity (\eg ``If I use chatbots, I think I would be able to depend on it completely'') to measure trust. 
The above items were evaluated using a five-point Likert scale (1 = Not at all, 5 = Strongly agree).

\subsection{Pilot Study}

Initially, we used the scale developed by Bennion and Adams~\cite{bennion1986revision} to measure ego identity. This scale covers eight core domains—four ideological and four interpersonal—each with eight items corresponding to four identity statuses (identity achievement, moratorium, diffusion, and foreclosure), yielding a total of 64 items. 
The remaining variables were measured using the aforementioned validated scales. In total, in the pilot study, the questionnaire contains 102 items.

We recruited 14 teenagers aged 16–18 (10 female, 4 male) to participate in the pilot study. After they finished, we conducted brief follow-up interviews, asking questions such as: ``How did you feel about the overall process of filling out this questionnaire?'' ``What do you think about the length of the questionnaire?'' and ``Were there any items that you found difficult to understand or were unsure about what they were asking?''

The interview results indicated that the 102-item questionnaire was overwhelming for most participants. Moreover, several participants reported difficulty responding to certain items on the ego identity scale—particularly those concerning religion and dating—due to their limited relevance and practical experience in the Chinese context. 
In other words, while the scale developed by Bennion and Adams~\cite{bennion1986revision} provides a comprehensive operationalization of the psychological theory of ego identity, its complexity and certain culture-specific items rendered it less suitable for the current research.
Therefore, the original ego identity measurement was substituted with a more concise alternative, namely the 12-item Ego Identity Scale by Tan et al.~\cite{tan1977short}. After this refinement, our questionnaire consisted of 50 items in total.

Next, we conducted a second round of pilot study, inviting another 37 teenagers (32 female, 5 male) aged 13–18 to complete the questionnaire and carrying out the same set of follow-up interviews. Results indicated that the 50-item questionnaire became acceptable to most teenagers, while a few participants reported impatience and loss of attention. Therefore, based on the empirical findings of Kung et al.~\cite{kung2018attention}, we included two attention-check questions in the questionnaire to help screen low-quality responses. 
Additionally, in the interviews, two participants told us that their ratings were swayed by imagined future improvements to chatbots. To counter this, in the informed consent form at the beginning of the questionnaire, we explicitly instructed participants to base their responses on their current experiences with chatbots rather than making assumptions or speculations about the future development of chatbot technology.


After the two rounds of revisions, we conducted a third round of pilot testing, inviting another 35 teenagers  (23 female, 12 male) aged 14–18 to participate in the survey. The performance of the questionnaire in this round was satisfactory, establishing the readiness for larger-scale data collection and laying the groundwork for the formal study.


\subsection{Formal Study}

Below, we introduce our formal study and findings.

\subsubsection{Participants and Procedure}

Following the definition proposed by the World Health Organization, we define teenagers as individuals aged 10-19 years~\cite{timmurphy.org}. 
As this age range corresponds to the junior and senior high school educational stages in China, we selected these schools as our recruitment sites and distributed 213 online questionnaires to students at one junior high school and two high schools in Shanghai, China.

Shanghai was selected as the primary site for data collection for two reasons. First, as a first-tier city in China, Shanghai is at the forefront of generative AI development. According to the 55th Statistical Report on China’s Internet Development~\cite{CNNIC2025}, Shanghai accounts for 19.9\% of all registered generative AI products nationwide, giving its residents greater exposure to chatbots and similar AI applications. This high penetration rate facilitated the identification and recruitment of teenagers with prior experience using AI chatbots. 
Second, pre-existing relationships between the research team and educational institutions in Shanghai allowed for smoother recruitment and greater cooperation from the schools involved. Drawing on our personal social networks, we established contact with four secondary schools in Shanghai. The researchers provided the teachers responsible for relevant tasks at each school with a detailed explanation of the study’s background, core objectives, and practical significance. One high school, due to internal management regulations, declined to participate in the questionnaire distribution. The remaining three schools acknowledged the scientific rigor and validity of the project and ultimately formed a cooperative partnership based on mutual trust, laying a solid foundation for the subsequent data collection process.

\x{According to our institution's policies at the time of the study, this type of research did not require a full ethical review by a formal Institutional Review Board (IRB). Nonetheless, the research team conducted an internal review adhering to the ethical principles of voluntary participation and informed consent. Given that all participants were minors, the researchers first explained the purpose of the study to schoolteachers. The teachers then distributed the voluntary survey to parents through the class parent groups, and only those students whose parents provided electronic informed consent were invited to participate.}
We used the online survey platform Wenjuanxing—a widely used Chinese tool for online questionnaire design and data collection—to develop the questionnaire and collect responses. Before starting the survey, participants were presented with an informed-consent form outlining the study’s background and purpose. The form emphasized that participation was anonymous, that collected data would be used solely for scientific research, and that no information would be publicly disclosed. Participants were also informed that they could withdraw at any time without consequence. Additionally, the questionnaire included an optional final item asking participants if they were willing to provide contact information for a potential follow-up interview.

All teenagers participated voluntarily and spent approximately 5-8 minutes completing the online questionnaire. 
After completing the survey, we conducted a rigorous screening process to exclude irrelevant and low-quality answers. First, we excluded responses in which participants answered ``No'' to the question ``Have you ever used AI chatbots (\eg ChatGPT, DeepSeek, Qwen, Claude, Doubao)?'' since teenagers who had never used chatbots were not the target population of this study. Second, we removed invalid questionnaires that failed the attention-check items or displayed unreasonable response patterns. Specifically, if a participant failed to select the required option on either of the two attention-check questions (\eg ``Please select `Strongly Agree' for this item'') or if their responses showed patterns such as repeatedly selecting the same option across nearly all the items, the questionnaire was deemed low-quality and excluded.

After excluding 61 questionnaires, we obtained 152 valid responses. Among these, 55 respondents were male and 97 were female; the mean age was 15.95 years; and the average completion time was 409 seconds. The 152 responses were used for the following data analysis.

\subsubsection{Data Analysis Methods}

\begin{figure}[htbp]
\centering
    \subcaptionbox{Q-Q Plot of AI Literacy\label{fig:AI Literacy}}
{\includegraphics[width=0.3\textwidth]{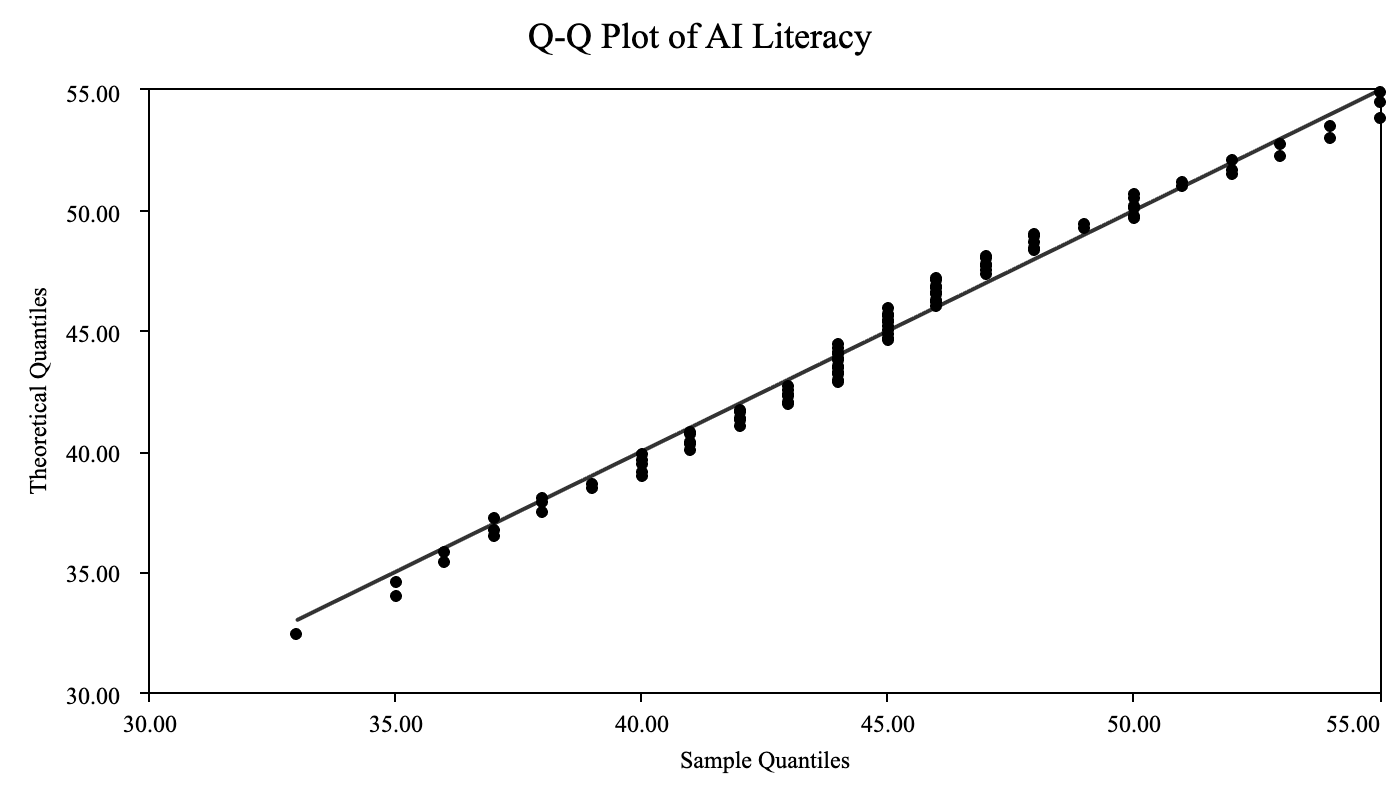}}
    \subcaptionbox{Q-Q Plot of Ego Identity\label{fig:Ego Identity}}
{\includegraphics[width=0.3\textwidth]{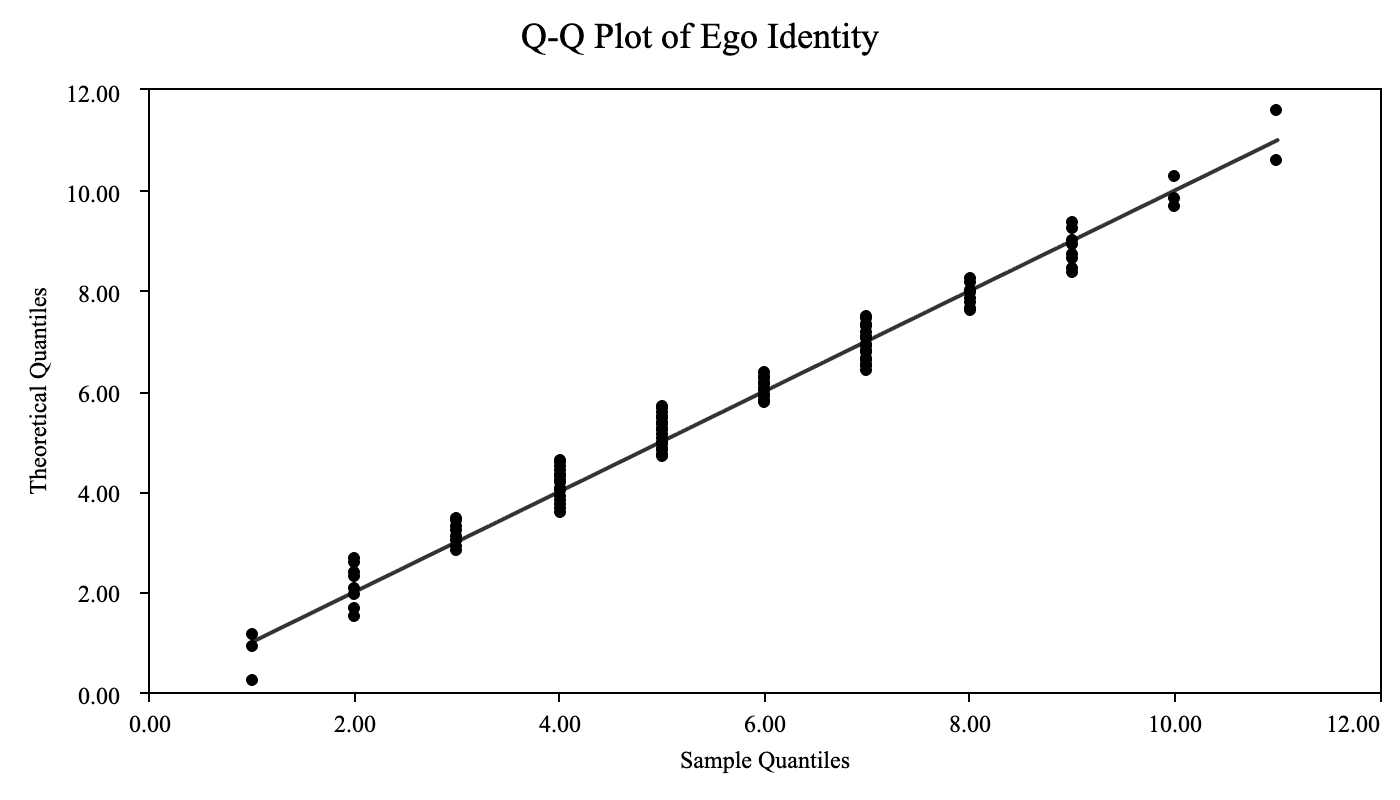}}
    \subcaptionbox{Q-Q Plot of Psychological Resilience\label{fig:Psychological Resilience}}
{\includegraphics[width=0.3\textwidth]{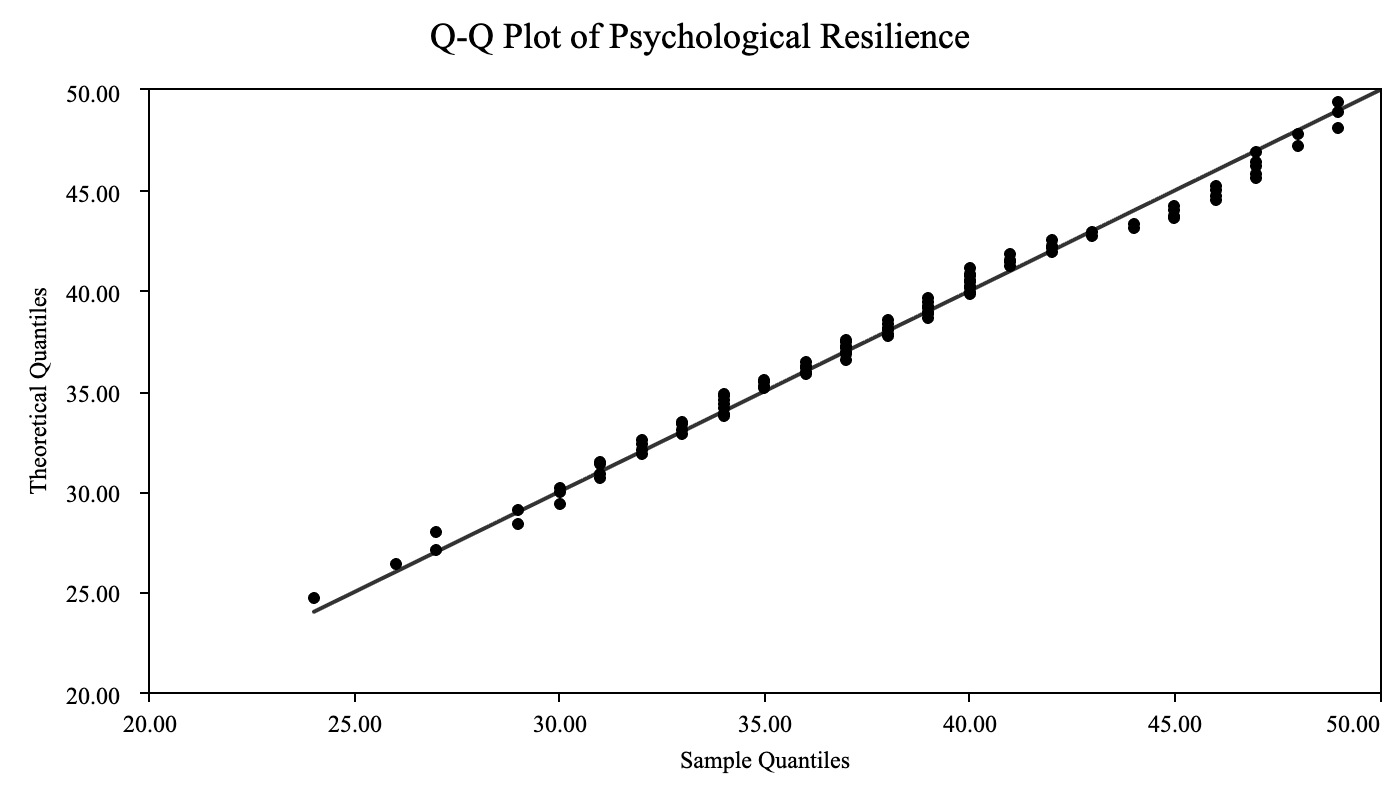}}
    \subcaptionbox{Q-Q Plot of Social Anxiety\label{fig:Social Anxiety}}
{\includegraphics[width=0.3\textwidth]{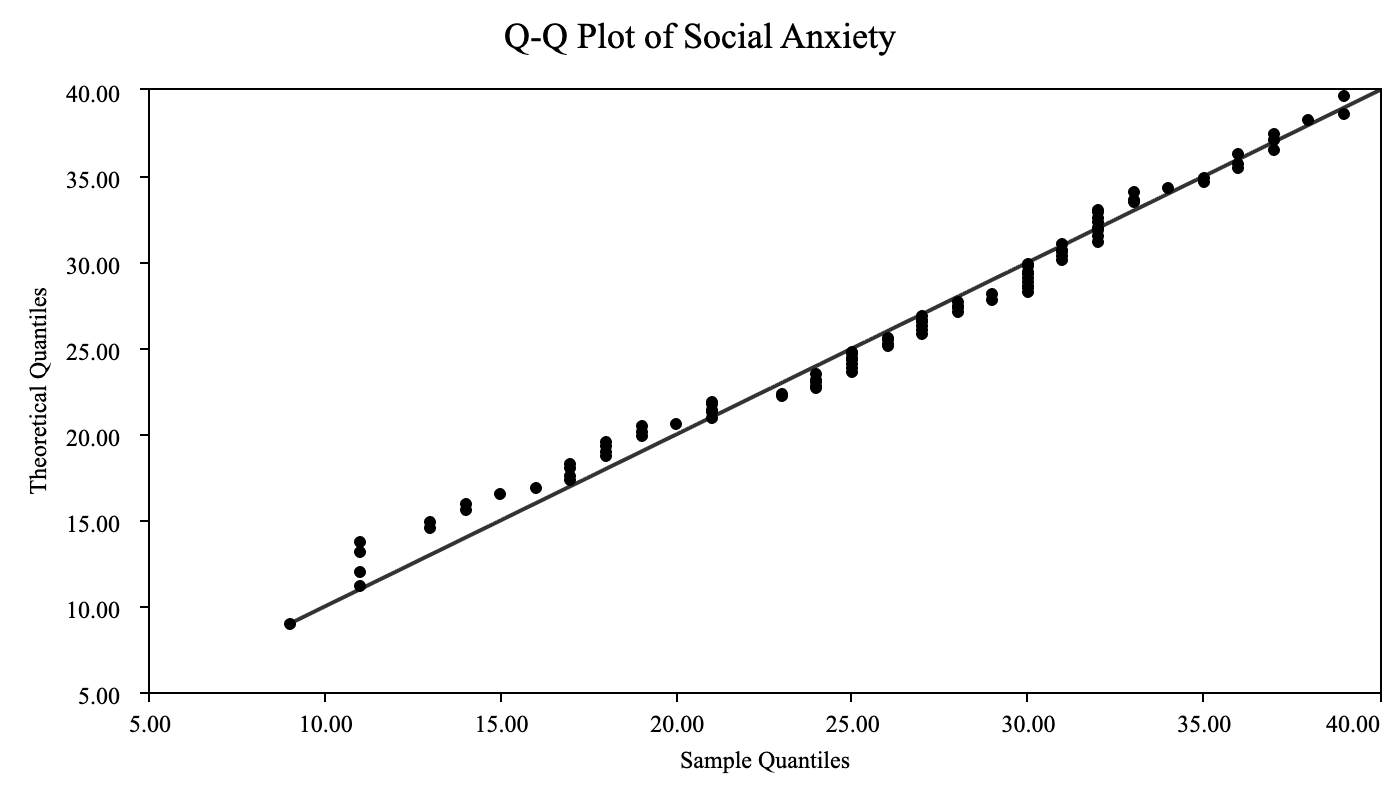}}
    \subcaptionbox{Q-Q Plot of Trust\label{fig:Trust}}
{\includegraphics[width=0.3\textwidth]{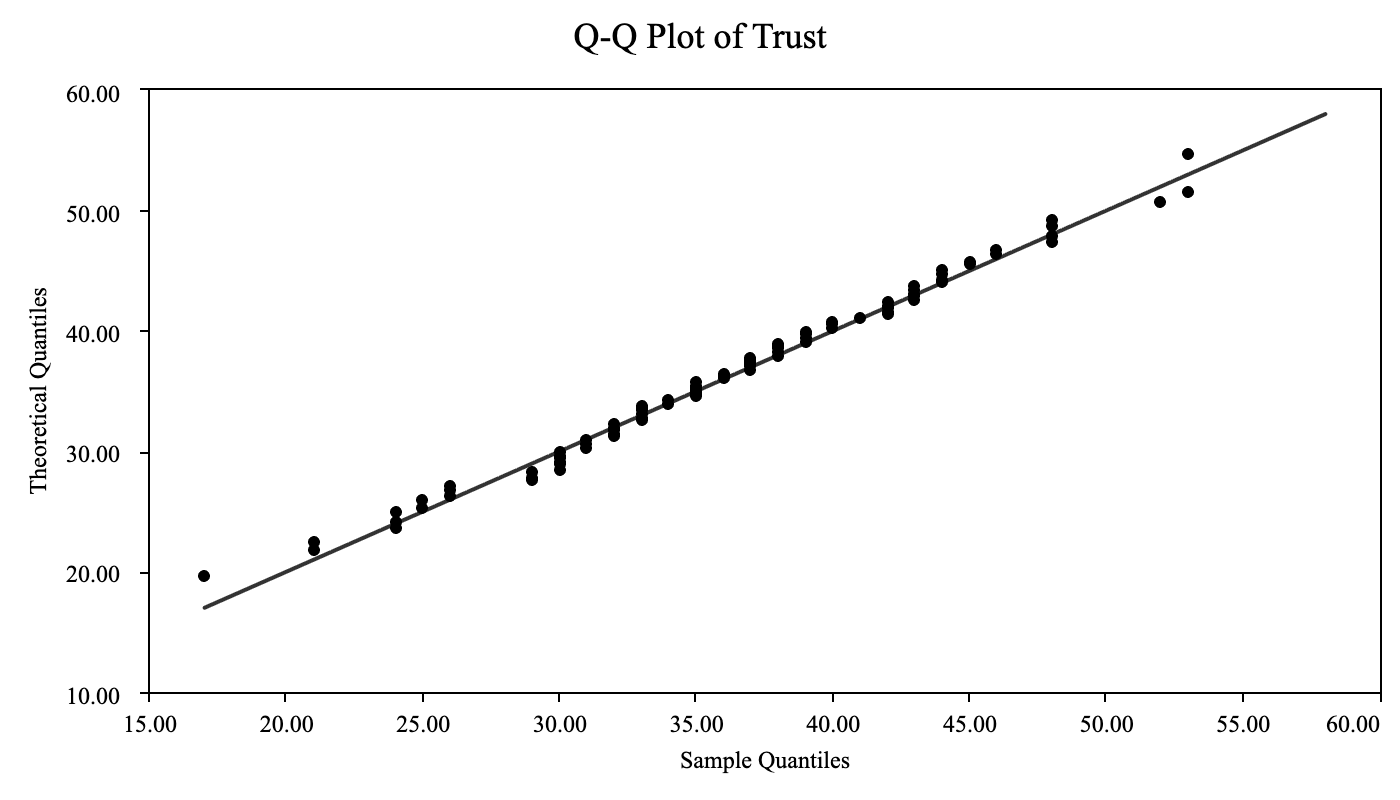}}
    \caption{Q-Q plots of tested variables}
    \label{fig:qq}
\end{figure}

The raw data were exported and analyzed using SPSS software. We first conducted internal consistency reliability analysis for the scale-type items using Cronbach’s $\alpha$ coefficient. The results showed that the Cronbach’s $\alpha$ coefficients for AI literacy, social anxiety, psychological resilience, and trust were 0.836, 0.902, 0.858, and 0.793, respectively, all indicating high reliability. After ensuring data validity, we used Q-Q plots to test the normality of the data and confirmed that they followed a normal distribution (see \autoref{fig:qq}). Subsequently, we performed reverse scoring for negatively worded items and assigned scores for single-choice questions to calculate the total score of each variable for each participant. Finally, Pearson correlation analysis was employed to examine the relationships between the independent variables and the dependent variable, with statistical significance set at $p < 0.05$.

\subsubsection{Analysis Results}

We conducted correlation analyses to examine the relationships among AI literacy, ego identity, social anxiety, psychological resilience, and trust in AI chatbots. \autoref{tab:kmo} presents the correlation coefficients and significance levels.

\begin{table}[t]
    \centering
    \fontsize{7.8}{9.5}\selectfont
    \caption{Correlation Coefficients of Study Variables}
    \label{tab:kmo}
    \begin{tabular*}{\columnwidth}{@{\extracolsep{\fill}}lc}
        \toprule
        \textbf{Variable} & \textbf{Trust} \\
        \midrule
        AI Literacy & 0.100 \\
        Ego Identity & 0.1301 \\
        Social Anxiety & 0.035 \\
        Psychological Resilience & 0.183* \\
        \bottomrule
    \end{tabular*}
    \vspace{2pt}
    \raggedright
    \fontsize{6.5}{8}\selectfont
    \textit{Note.} * $p<.05$; ** $p<.01$.
\end{table}

\paragraph{Psychological Resilience and Trust in AI}

We observed a significant positive correlation between psychological resilience and trust in AI chatbots ($r = 0.183, p < 0.05$). Thus, \textbf{H4} was supported.

\paragraph{AI Literacy and Trust in AI}

As shown in Table 1, the direction of the correlation between AI literacy and trust in AI aligns with our hypothesis; however, the relationship did not reach the predetermined level of statistical significance ($r = 0.10, p = 0.222$). This indicates that the current data do not provide sufficient evidence of a significant linear relationship between AI literacy and trust in AI. Therefore, \textbf{H1} was not supported.

\paragraph{Ego Identity and Trust in AI}
Similarly, the correlation between ego identity and trust in AI was in the hypothesized direction but failed to reach statistical significance ($r = 0.13, p = 0.112$). Consequently, the data do not offer sufficient evidence for a significant linear relationship between ego identity and trust in AI, and \textbf{H2} was not supported.

\paragraph{Social anxiety and Trust in AI}

The correlation between social anxiety and trust in AI was in the hypothesized direction but did not reach the predetermined level of statistical significance ($r = -0.035, p = 0.668$). This indicates that the current data provide insufficient evidence for a significant linear relationship between social anxiety and trust in AI; therefore, \textbf{H3} was not supported.

Analysis of the questionnaire data revealed generally weak correlations among the core variables. To further examine potential factors influencing the relationships between independent and dependent variables, we introduced demographic variables—specifically age and gender—for additional analysis to determine whether these characteristics moderated the observed relationships.

The study found that gender did not have a moderating effect but exerted a significant direct effect on trust. Specifically, there was a statistically significant difference in trust across physiological gender, with male teenagers generally exhibiting higher levels of trust in AI compared to female teenagers (see \autoref{tab:anova_gender_trust}).

\begin{table}[t]
    \caption{ANOVA Results for Gender Differences in Trust}
    \label{tab:anova_gender_trust}
    \centering
    \fontsize{7.8}{9.5}\selectfont
    \begin{tabularx}{\columnwidth}{lcccc}
        \toprule
        \multirow{2}{*}{} &
        \multicolumn{2}{c}{\textbf{Physiological Gender (Mean ± SD)}} &
        \multirow{2}{*}{\textbf{F}} &
        \multirow{2}{*}{\textbf{p}} \\
        \cmidrule(lr){2-3}
        & \textbf{Male (n=55)} & \textbf{Female (n=97)} &  &  \\
        \midrule
        Trust & 39.11 ± 7.69 & 34.52 ± 6.81 & 14.542 & 0.000** \\
        \bottomrule
    \end{tabularx}
    \vspace{2pt}
    \raggedright
    \fontsize{6.5}{8}\selectfont
    \textit{Note.} $n=152$. * $p<.05$; ** $p<.01$.
\end{table}

\begin{table*}[t]
    \centering
    \fontsize{7.8}{9.5}\selectfont
    \caption{Mediation Effect Models for Role 1}
    \label{tab:role1}
    \begin{tabular*}{\textwidth}{@{\extracolsep{\fill}}lcccccccccccc@{}}
        \toprule
        & \multicolumn{4}{c}{Model 1} 
        & \multicolumn{4}{c}{Model 2} 
        & \multicolumn{4}{c}{Model 3} \\
        \cmidrule(lr){2-5}
        \cmidrule(lr){6-9}
        \cmidrule(lr){10-13}
        & \textbf{B} & \textbf{t} & \textbf{p} & \textbf{$\beta$}
        & \textbf{B} & \textbf{t} & \textbf{p} & \textbf{$\beta$}
        & \textbf{B} & \textbf{t} & \textbf{p} & \textbf{$\beta$} \\
        \midrule
        Constant
        & 36.178 & 59.710 & $<.001^{**}$ & --
        & 36.178 & 59.722 & $<.001^{**}$ & --
        & 36.168 & 61.012 & $<.001^{**}$ & -- \\
        Social Anxiety
        & -0.033 & -0.429 & 0.668 & -0.035
        & -0.034 & -0.437 & 0.663 & -0.036
        & -0.020 & -0.268 & 0.789 & -0.021 \\
        Age
        &  &  &  &
        & -0.474 & -1.030 & 0.304 & -0.084
        & -0.840 & -1.790 & 0.076 & -0.149 \\
        Social Anxiety $\times$ Age
        &  &  &  &
        &  &  &  &
        & -0.137 & -2.756 & 0.007$^{**}$ & -0.230 \\
        \midrule
        $R^2$
        & \multicolumn{4}{c}{0.001}
        & \multicolumn{4}{c}{0.008}
        & \multicolumn{4}{c}{0.057} \\
        Adjusted $R^2$
        & \multicolumn{4}{c}{-0.005}
        & \multicolumn{4}{c}{-0.005}
        & \multicolumn{4}{c}{0.038} \\
        \textit{F}
        & \multicolumn{4}{c}{$F(1,150)=0.184,\,p=0.668$}
        & \multicolumn{4}{c}{$F(2,149)=0.623,\,p=0.538$}
        & \multicolumn{4}{c}{$F(3,148)=2.966,\,p=0.034$} \\
        \bottomrule
    \end{tabular*}
    \vspace{3pt}
    \par
    \raggedright
    \fontsize{6.5}{8}\selectfont
    \textit{Note.} $n=152$. * $p<.05$; ** $p<.01$.
\end{table*}

\begin{figure}[h]
\centering
\includegraphics[width=0.7\columnwidth]{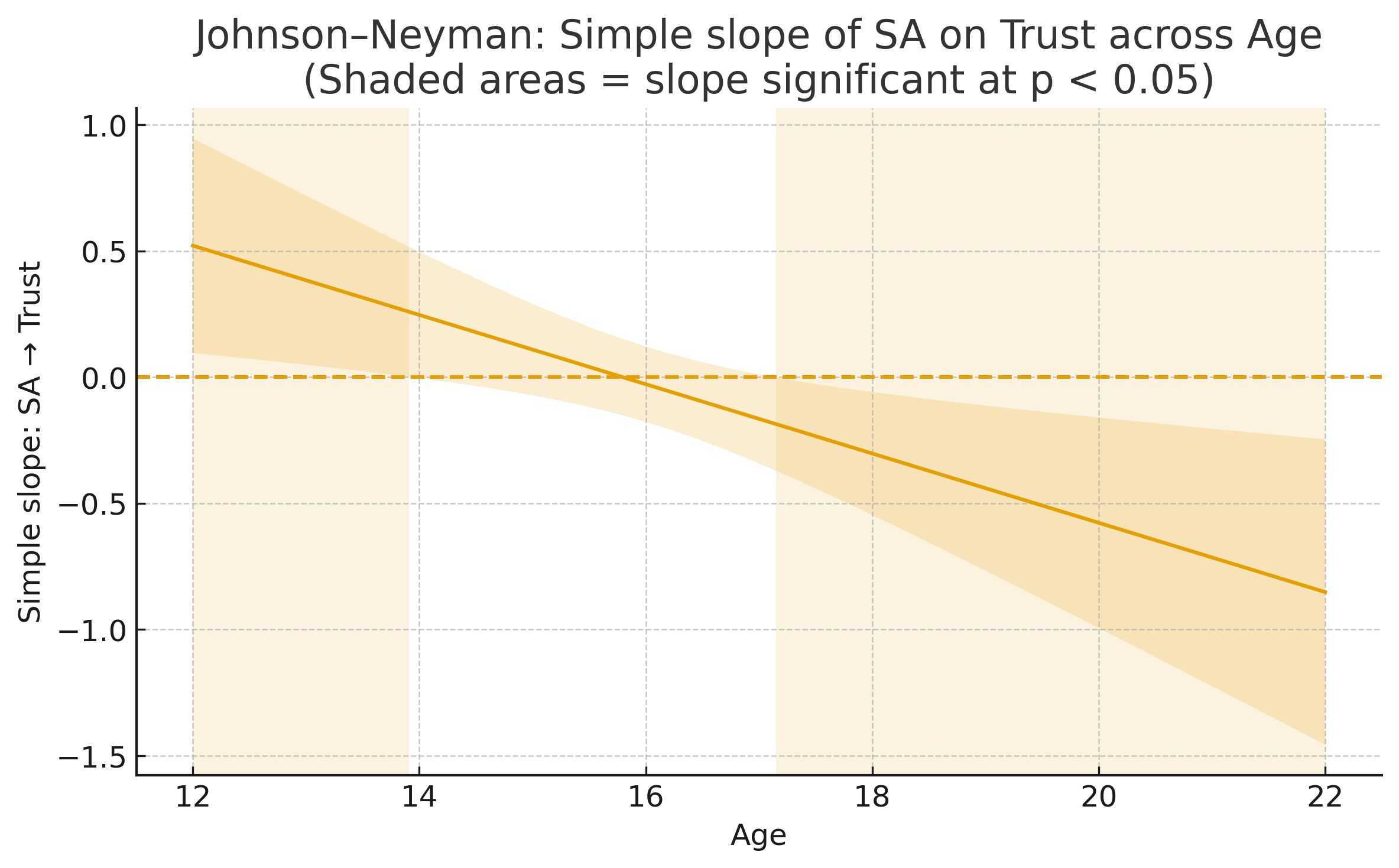}
\caption{Johnson–Neyman plot illustrating the simple slope of social anxiety predicting trust across age}
\label{fig:Johnson–Neyman plot}
\end{figure}

\begin{figure}[h]
\centering
\includegraphics[width=0.7\columnwidth]{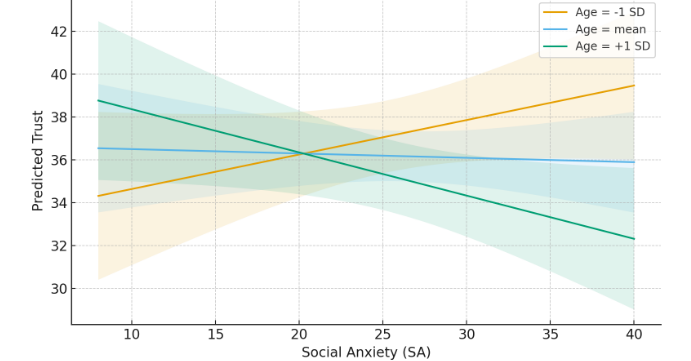}
\caption{Predicted trust as a function of social anxiety at three age levels (-1 SD, mean, +1 SD)}
\label{fig:Age}
\end{figure}

The results also showed that age significantly moderated the effect of social anxiety on trust: among older teenagers, higher levels of social anxiety were associated with a significant decrease in trust, whereas among younger teenagers, social anxiety had no significant impact on trust (see \autoref{tab:role1}).
The simple slope analysis of this moderating effect using the Johnson-Neyman method is shown in \autoref{fig:Johnson–Neyman plot}. It indicated that at lower age levels (–1 standard deviation), the association between social anxiety and trust was positive but not statistically significant ($b = 0.161, SE = 0.104, t = 1.555, p = 0.122, 95\% CI = [-0.044, 0.366]$). At the mean age level, the association was nearly zero and also not significant ($b = -0.020, SE = 0.076, t = -0.27, p = 0.789, 95\% CI = [-0.170, 0.130]$). In contrast, at higher age levels (+1 standard deviation), the negative effect of social anxiety on trust reached statistical significance ($b = -0.202, SE = 0.097, t = -2.08, p = 0.040, 95\% CI = [-0.394, -0.010]$).

Further Johnson–Neyman analysis revealed that the negative effect of social anxiety on trust becomes significant when age exceeds \x{16.52} years. This suggests that social anxiety primarily undermines AI trust among older teenagers, whereas its impact is negligible for younger and average-age participants (see \autoref{fig:Age}). These findings highlight an age-related vulnerability in the relationship between social anxiety and trust in AI.

\textbf{In summary,} through this questionnaire survey, although the directions of influence of the independent variables on the dependent variable were consistent with the hypotheses, only psychological resilience exhibited a statistically significant effect. The effects of AI literacy, ego identity, and social anxiety were not statistically significant. Moreover, we found that gender had a significant impact on trust, and age significantly moderated the relationship between social anxiety and trust.

\section{Semi-Structured Interviews}
\label{sec:interview}

Since the statistical results of the questionnaire did not fully support our hypotheses, we conducted additional interviews to explain why certain ratings deviated from the model’s predictions.

\subsection{Participants}
From the valid questionnaires, we selected 15 individuals with extensive experience using AI chatbots who were also willing to be interviewed. \autoref{tab:Participant Information.} presents information about the 15 participants, including 3 males and 12 females, with ages ranging from 16 to 19 years old ($M = 17.67, SD = 0.82$). Their duration of chatbot use varied from less than one month to over a year, and DeepSeek and ChatGPT are their most used chatbots.

\begin{table}[t]
    \centering
    \fontsize{7.8}{9}\selectfont
    \caption{Participant Information (M denotes Male, and F denotes Female).}
    \label{tab:Participant Information.}
    \begin{tabular*}{\columnwidth}
    {p{1em}p{1em}p{1em}p{7em}l}
        \toprule
        \textbf{No.} 
        & \textbf{Age} 
        & \textbf{Sex} 
        & \textbf{Duration of Chatbot Usage} 
        & \textbf{Chatbot(s) Used} \\
        \midrule
        1  & 17 & M   & 3--6 months      
           & DeepSeek, ERNIE Bot, Doubao \\
        2  & 16 & F & 3--6 months      
           & ChatGPT, DeepSeek, Doubao \\
        3  & 18 & F & > 1 year      
           & DeepSeek, bimobimo \\
        4  & 18 & F & 6 months--1 year 
           & DeepSeek, Xiaoyi \\
        5  & 16 & F & < 1 month
           & DeepSeek \\
        6  & 18 & F & 6 months--1 year 
           & ChatGPT, DeepSeek \\
        7  & 18 & F & > 1 year      
           & DeepSeek, ERNIE Bot, Doubao \\
        8  & 18 & M   & > 1 year      
           & ChatGPT, DeepSeek, Claude, Gemini, Doubao \\
        9  & 19 & F & > 1 year      
           & ChatGPT, DeepSeek \\
        10 & 17 & F & 6 months--1 year 
           & ChatGPT, DeepSeek, Gemini, Doubao \\
        11 & 18 & F & 6 months--1 year 
           & ChatGPT, Kimi, Doubao \\
        12 & 18 & F & 1--3 months      
           & DeepSeek \\
        \bottomrule
    \end{tabular*}
    \vspace{3pt}
    \raggedright
    \fontsize{6.5}{8}\selectfont
\end{table}

\subsection{Methodology}

The interviews were conducted individually through online conferencing software in a semi-structured format. Before the interview, we explained the main objectives of the study, assured participants that the results would be used solely for academic purposes, and obtained their consent for audio recording. Since the purpose of the interviews was to better validate the findings from the questionnaire survey, the interview questions focused mainly on participants’ specific chatbot usage scenarios, methods, and the extent to which different variables influenced teenagers’ trust in AI, as well as the reasons behind such influences. This provided qualitative explanations for the quantitative results.
Typical interview questions we asked included: ``Why did you set a high score to the trust question?'', ``Under certain circumstances, would you prefer to defer judgment to a chatbot rather than relying on your own? '', ``Would a lack of understanding about how chatbots work lead you to trust them less? '', ``When facing interpersonal difficulties—such as fear of rejection or hesitation to speak up—would you turn to a chatbot for advice?'', ``Are there any external factors that influence your level of trust in chatbots?''

After each interview, recordings were transcribed into text. The interview transcripts were imported into NVivo for manual coding. We employed a bottom-up, open coding process to analyze the underlying reasons for the three variables' lack of significant impact on teenagers' trust in AI.
As a result, we summarized and categorized relevant insights into three main themes: (i) Trust remains consistently high regardless of variations in psychological factors, (ii) Overestimated AI literacy and self-report biases, (iii) The potential influence from social media.

\subsection{Results}

Below, we present a detailed analysis of the interview findings, which offer critical context for our quantitative results.

\subsubsection{R1: Trust remains consistently high regardless of variations in psychological factors.} 

By closely examining their raw ratings scores as well as conducting interviews, we noticed that the majority of teenagers demonstrated a relatively high level of trust in AI. 

\paragraph{R1.1: The fact that learning constitutes the dominant context for Chinese teenagers' AI use skewed their perception of AI.}
A predominant theme emerging from the interviews was the instrumental and domain-specific nature of chatbot use, which was overwhelmingly concentrated within learning contexts. Participants consistently reported leveraging chatbots as high-efficiency tools to support their schoolwork. For instance, they used them for debugging code (Participant 2), completing assignments in the humanities (Participant 3), and collecting source materials for writing projects (Participant 4). Within this functional domain, teenagers expressed a high degree of trust, which was rooted in a rational assessment of the chatbots’ superior capabilities and efficiency compared to their own. This pragmatic trust was articulated clearly by Participant 5, who stated: ``\textit{If I’m unsure about a question, choose option B, and the chatbot tells me the answer is D, I’ll definitely believe it.}'' This highlights a form of cognitive trust based purely on perceived competence.

Besides, our data indicated that while some participants used chatbots for emotional or recreational purposes—scenarios less conducive to building strong trust—such uses were significantly less common than learning-related interactions. For example, several participants were sensitive to anthropomorphic framing. When asked, ``Do you think the chatbot tries its best to help you?'' Participant 6 emphasized, ``\textit{The phrase 'tries its best' sounds too anthropomorphic. I think it's more accurate to interpret it in terms of effectiveness.}''

Several factors explain this tendency: (1) Teenagers perceived chatbot responses as ``excessively compliant.'' As Participant 8 noted, ``\textit{Conflict is a crucial part of emotional experience. Unconditional compliance only offers short-term, temporary, and unreliable emotional value,}'' thus failing to meet deeper emotional needs. (2) Although some introverted participants with mild social anxiety found interacting with chatbots easier than communicating in real-life settings—suggesting that social anxiety may indirectly increase AI trust in functional contexts by reducing perceived social pressure—those who engaged with chatbots also reported that such interactions remained superficial. For example, when facing genuine emotional difficulties, they would not seek support from chatbots (Participant 3). This indicates that the relationship between social anxiety and AI trust is likely context-dependent. (3) Many teenagers approached chatbots with the preconception that AI is objective and emotionless, which shaped an interaction style that consciously avoided emotional content. As Participant 6 explained, ``\textit{I feel that AI inherently gives an impression of being rational, stable, and trustworthy. My trust doesn’t come from how long it has accompanied me or how much emotional value it provides, but rather from the sense that there’s always someone—or something—that will respond when I need help in daily life.}'' (4) Those with higher self-identity preferred to face challenges independently rather than seeking emotional reliance on chatbots. Participant 4 stated, ``\textit{I believe solving problems should be done by oneself. If I always rely on a machine for support, I’ll never become someone who can handle situations well. That kind of dependence isn’t good for my growth.}''

On the other hand, the deep-rooted cultural emphasis on academic prioritization in Chinese society reinforces a purely functional perception of AI. Most interviewees indicated that while teachers and parents encouraged them to use AI for learning support, guidance regarding emotional or relational applications was virtually absent. At the same time, some respondents reported strict parental limits on daily electronic device usage, which further restricted their exposure to AI. As a result, teenagers tend to consciously avoid emotional interactions with AI, prioritizing academically oriented functions instead—thereby diminishing the impact of psychological traits on trust formation.

This narrowly defined, learning-centric mode of interaction likely skewed participants' interpretations of the questionnaire items, causing them to evaluate trust primarily from a functional, academic perspective rather than a broader psychological one. This limited scope of engagement may explain the weakened correlations observed in our quantitative data between psychological traits (like self-identity and social anxiety) and a generalized measure of trust. If chatbots are perceived merely as tools for schoolwork, their use may not activate the deeper psychological dispositions that govern interpersonal relationships, thus rendering those traits less relevant to the formation of trust in this specific context.

\paragraph{R1.2: High tolerance of unsatisfactory responses.}
Interview data revealed that 14 out of the 15 participants explicitly mentioned having experienced unsatisfactory responses from AI chatbots, including instances of fabricated information, logical inconsistencies, and misinterpretation of requests. Some of these negative experiences directly affected core usage needs such as learning tasks or important information retrieval. For example, Participant 6 noted that chatbots sometimes generated fictional references when assisting with essay writing, while Participant 8 found that queries related to less common topics in Chinese literature, history, and philosophy often received distorted or fabricated answers. Similar issues appeared in everyday scenarios, where the outputs were inconsistent or impractical, such as when a participant asked the AI to reproduce the speech patterns of a novel character and found the result confused and mismatched with the original details. In more emotionally oriented scenarios, participants also expressed disappointment at the lack of empathy, as illustrated by a 19-year-old female participant who reported that when she confided interpersonal difficulties to the AI, its responses were formulaic and detached from her personal situation.

Despite these shortcomings, the majority of participants rarely chose to abandon the tool entirely. As Participant 6 explained, ``\textit{I don’t completely distrust it, nor would I stop using it because of occasional errors. If it gives wrong answers repeatedly, I’ll leave the current chat window and start a new one with rephrased input.}'' Instead, they tended to respond with correction, avoidance, or tolerance. Many attributed incorrect answers to their own use methods rather than to the chatbot’s unreliability and attempted to refine their prompts to improve outputs. Even when encountering a series of unsatisfactory responses, participants were more likely to reduce the frequency of asking similar questions in the future rather than generalizing distrust to the specific chatbot product or the technology as a whole.

\subsubsection{R2: Overestimated AI literacy and self-report biases.}

In the interviews, most teenagers reported having a relatively good understanding of AI. However, when interviewers probed further about AI’s core operational logic, participants were unable to provide specific or systematic explanations and instead described only surface-level functions such as ``\textit{it can chat}'' or ``\textit{it can answer questions.}'' Participant 6 commented, ``\textit{If one day I could understand its logic more deeply, I might trust it more.'}'

This suggests that teenagers may exhibit self-report bias regarding AI literacy on questionnaires—overestimating their literacy levels based on superficial exposure to AI—which leads to inflated scores on the AI literacy dimension and masks the true relationship between AI literacy and AI trust. The underlying reason for this superficial perception is likely that within the Chinese secondary education context, teenagers' AI use is largely confined to academic support scenarios (e.g., doing homework, organizing materials). These scenarios demand only basic operational skills from them, without requiring a deep understanding of AI's principles and ethical risks.

\subsubsection{R3: The potential influence from social media}
The interviews also revealed that the external information environment—particularly social media—may act as a potential moderating variable influencing teenagers' trust in AI. Fragmented narratives about AI on social media platforms (e.g., ``AI diminishes independent thinking,'' ``AI will replace humans'') can disrupt teenagers’ trust in AI. \x{For example, Participant 5 said, ``\textit{I often come across videos that talk about 'robots defeating humans' or similar topics. It makes me think that as AI development continues to grow, many of our job opportunities may decrease. Moreover, if I share a lot about myself with AI and let it understand me well enough, it might eventually influence my emotions. That's why I try not to talk about these things.}''} Since teenagers’ cognitive judgment is still developing, exposure to such external narratives may lead them to modify their frequency of chatbot use or avoid certain usage scenarios, thereby further weakening the relationship between psychological variables and AI trust. 


\x{To sum up, by combining the findings from the questionnaires and interviews, we found that although most surveyed teenagers gave chatbots relatively high trust ratings, their engagement was primarily for functional and instrumental purposes. Their trust seems to stem from perceiving AI as a highly effective tool, despite its flaws, and may be facilitated by a limited grasp of the technology. However, we did not observe evidence of emotional overreliance in their daily use, and therefore, they did not reach the extreme levels of dependence reported in some media accounts.}
\section{Discussion}

Below, we discuss our major findings and limitations.

\subsection{The Positive Effect of Psychological Resilience on AI Trust}

This study revealed a significant positive correlation between teenagers' psychological resilience and their trust in AI chatbots. This finding was also supported by interview data. For instance, Participants 8 and 15, who exhibited higher resilience, maintained independent judgment in critical decisions (e.g., choosing research topics or planning their future), which corresponded with a stronger sense of autonomy and a higher level of trust in AI. In contrast, Participants 9 and 12, with lower self-reported resilience, expressed greater ambivalence toward AI, which ultimately weakened their trust and led them to prefer self-adjustment or human support.

\x{These findings suggest that integrating resilience training into daily teaching and mental health education may help teenagers cope more effectively with stress and negative emotions. This, in turn, would support them in maintaining rational judgment when using AI. To cultivate such resilience, a multi-faceted approach is recommended. This includes fostering individual competencies through practices like mindfulness and physical exercise~\cite{mei2025interventions}, and strengthening social support, with a particular emphasis on the intimacy derived from close bonds~\cite{monasterio2002enhancing}.} Moreover, \x{courses on stress management and psychological education can be integrated into school curricula~\cite{kallianta2021stress,deckro2002evaluation}.} Such practices may strengthen resilience and guide teenagers to develop a more balanced trust in AI, avoiding the risks of misuse caused by insufficient psychological strengths.

\subsection{The Significant Moderating Role of Age in the Relationship Between Social Anxiety and AI Trust}

This study further examined the moderating role of demographic variables and found that age significantly changes the effect of social anxiety on AI trust. When the age of a teenager exceeds 16.52 years, social anxiety is negatively correlated with AI trust ($b = –0.202, p = 0.040$). When the age is below 16.52 years, social anxiety shows no significant effect ($b = 0.161, p = 0.122$). In other words, among older teenagers, the higher the level of social anxiety, the lower the level of trust in AI. This may also explain why the overall correlation analysis did not show a significant negative effect. This moderating effect aligns with previous literature, such as research examining whether teenagers become more sensitive to social evaluation as they age~\cite{sumter2010age}, which indicates that biological stress sensitivity increases during adolescence, at least in response to a social-evaluative situation.

This result also challenges studies that tended to treat teenagers as a homogeneous group and overlooked age-based differences~\cite{piombo2025emotional,lee2025understanding}. It highlights 16.52 years as a potential threshold at which social anxiety begins to exert a significantly negative influence on AI trust. In the Chinese context, this age group typically corresponds to the final two years of high school (grades 11-12). During this period, most students face immense pressure preparing for the National College Entrance Examination (Gaokao), which can lead to significant conflicts with their parents~\cite{cao2023patterns}. Their social status was correlated mainly with intelligence but also with physical attractiveness~\cite{Qi1996correlates}, creating an environment of potential psychological and social challenges.

Together, these findings suggest that AI design and educational guidance should consider the age-specific needs of teenagers. For those aged 17 and above, AI systems should strengthen transparency and explainability in order to reduce the sense of uncontrollability that may amplify social anxiety. 

\subsection{Limitations and Future Directions}
This study has several limitations. First, recruiting minors for research purposes is challenging due to the need to obtain both parental and school consent. This difficulty leads to a relatively small sample size; future research should consider expanding the sample to improve representativeness and generalizability. Second, while the sample of teenagers from Shanghai may be representative of early adopters of AI chatbot, differences in regional economic development and age distribution may limit the universality of the findings. Future studies could broaden the sampling scope to include more regions and a wider range of age groups, thereby enhancing the applicability of the results. Third, the reliance on questionnaires and semi-structured interviews as the main research methods may introduce subjective biases, which could affect the interpretation and generalization of the findings.

\x{From the perspective of interviews, another limitation lies in the insufficient representation of the diverse contexts in which teenagers use AI. Although some interviewees mentioned other AI use scenarios—such as casual conversation and emotional expression—the number of such cases was limited, learning-related contexts remain the dominant area of AI use among the studied teenagers. Thus, it is difficult to generalize these findings into broader conclusions. This points to a potential direction for future research: to explore non-academic scenarios of AI use, thereby enriching our understanding of teenagers’ patterns of engagement with AI.}

The researchers suggest that future studies on teenagers’ AI use and trust should further explore the human–AI relationship. For instance, whether such relationships evolve with technological updates, and whether teenagers’ judgments of AI’s anthropomorphic qualities determine the extent of AI’s impact on them. Such inquiries could yield deeper insights into teenagers’ patterns of media use in the age of artificial intelligence. \x{Additionally, the trust relationship between humans and AI should not be viewed as one-directional. While this study only considered teenagers’ trust in AI, the question of AI’s trust in humans is also a critical issue~\cite{Kashima2024trustworthy}. Future research could therefore investigate the bidirectional nature of trust between AI and teenagers, providing a more comprehensive understanding of mutual trust in human–AI interactions. Moreover, future work could further differentiate the specific qualities of trust itself. Beyond examining trust as a general attitude, researchers may distinguish between calibrated or appropriate trust, overtrust or misplaced trust, and undertrust~\cite{lee2004trust}. Such distinctions are crucial because different forms of trust may lead to divergent behavioral outcomes, such as inappropriate reliance, unwarranted skepticism, or balanced engagement. A more fine-grained conceptualization of trust would therefore deepen our understanding of how teenagers’ trust in chatbots translates into real-life decisions.}
\section{Conclusion}

This study investigates the relationships between teenagers' trust in AI chatbots and four factors: AI literacy, self-identity, social anxiety, and psychological resilience. Employing a mixed-methods approach, we found a significant positive correlation between psychological resilience and trust. Age significantly moderated the relationship between social anxiety and trust, with older teenagers exhibiting a stronger negative correlation.
To interpret the non-significant results for other variables, our qualitative analysis suggests that Chinese middle-school students currently use AI primarily as a learning tool, with minimal integration into their daily or emotional lives. This narrow scope of use, combined with potential self-report biases (such as overestimated AI literacy) and the influence of external factors like social media discourse, may have obscured the effects of variables like AI literacy and self-identity.

Practically, this study offers guidance for multiple stakeholders. Educators are encouraged to integrate AI literacy and psychological resilience training into curricula to help students develop healthy, critical trust in AI. Parents should focus on their children's psychological well-being and guide them toward balanced AI use. Social media platforms have a role in strengthening science communication to foster a rational public environment for AI adoption, supporting multi-dimensional and healthy teen-AI interactions.


\begin{acks}
This work was supported in part by the National Natural Science Foundation of China 62402121, Shanghai Chenguang Program, and Research and Innovation Projects from the School of Journalism at Fudan University.
\end{acks}

\bibliographystyle{ACM-Reference-Format}
\bibliography{sample-base}


\end{document}